\documentclass[aps,prd,superscriptaddress,twocolumn]{revtex4-1}

\usepackage{wallpaper}
\usepackage{bm}
\usepackage{amsmath}
\usepackage[colorlinks,linkcolor=blue,anchorcolor=blue,citecolor=red]{hyperref}

\begin{document}

\title{Compact dwarfs made of light-quark nuggets}
\author{Hao-Song You}
\affiliation{Center for Gravitation and Cosmology, College of Physical Science and Technology, Yangzhou University, Yangzhou 225009, China}

\author{Hao Sun}
\affiliation{Center for Gravitation and Cosmology, College of Physical Science and Technology, Yangzhou University, Yangzhou 225009, China}

\author{Hong-Bo Li}
\affiliation{School of Physics, Peking University, Beijing 100871, China}
\affiliation{Kavli Institute for Astronomy and Astrophysics, Peking University, Beijing 100871, China}


\author{Cheng-Jun Xia}
\email{cjxia@yzu.edu.cn}
\affiliation{Center for Gravitation and Cosmology, College of Physical Science and Technology, Yangzhou University, Yangzhou 225009, China}

\author{Ren-Xin Xu}
\affiliation{School of Physics, Peking University, Beijing 100871, China}
\affiliation{Kavli Institute for Astronomy and Astrophysics, Peking University, Beijing 100871, China}

\date{\today}

\begin{abstract}
Utilizing an equivparticle model with both linear confinement and leading-order perturbative interactions, we obtain systematically the properties of strangelets and nonstrange quark matter ($ud$QM) nuggets at various baryon ($A$) and charge ($Z$) numbers, where the detailed single-quark-energy levels are fixed by solving Dirac equations in mean-field approximation (MFA). We then examine the structures of compact dwarfs made of light strangelets or $ud$QM nuggets forming body-centered cubic lattices in a uniform electron background. Despite the strangelets and $ud$QM nuggets generally become more stable at larger $A$, the compact dwarfs are still stable since the fusion reactions between those objects do not take place in the presence of a Coulomb barrier, which is similar to the cases of light nuclei in normal white dwarfs. If $ud$QM dwarfs or strangelet dwarfs are covered with normal matter, their masses and radii become larger but do not exceed those of ordinary white dwarfs. Finally, we investigate the radial oscillation frequencies of $ud$QM dwarfs and strangelet dwarfs, and find that their frequencies are typically higher than traditional white dwarfs. The stability of compact dwarfs are then analysised  by examining  radial oscillation frequencies of the fundamental mode, where compact dwarfs covered by normal matter are still stable.
\end{abstract}


\maketitle

\section{\label{sec:intro}Introduction}
Stable nuclei in nature are expected to have a lifetime longer than the age of the Universe. Nevertheless, it is possible that nuclear matter may not be the true ground state of strongly interacting matter and could collapse into a new state of matter, i.e., quark matter~\cite{Bodmer1971_PRD4-1601}. It was postulated that strange quark matter (SQM) composed of approximately equal quantities of $u$, $d$, and $s$ quarks is more stable than nuclear matter~\cite{Witten1984_PRD30-272, Farhi1984_PRD30-2379, Terazawa1989_JPSJ58-3555}. This indicates the possible existence of stable SQM objects, e.g., strangelets ($A\lesssim10^7$)~\cite{Farhi1984_PRD30-2379, Berger1987_PRC35-213, Gilson1993_PRL71-332, Peng2006_PLB633-314, Xia2018_PRD98-034031, Chen2022_PRD105-014011, You2024_PRD109-034003}, nuclearites~\cite{Rujula1984_Nature312-5996, LOWDER1991_NPB24-177}, meteor-like compact ultradense objects~\cite{Rafelski2013_PRL110-111102} and strange stars ($A\approx 10^{57}$)~\cite{Itoh1970_PTP44-291,Alcock1986_APJ310-261}. The dark matter was speculated to be comprised of SQM objects existing before the nucleosynthesis era, offering an explanation within the framework of standard model~\cite{Zhitnitsky2003_JCAP2003-010, Banerjee2003_MNRAS340-284, Forbes2010_PRD82-083510}. Additionally, if the effects of spontaneous chiral symmetry breaking is considered, SQM becomes unstable due to a too large $s$ quark mass~\cite{Buballa1999_PLB457-261, Klahn2015_APJ810-134}, while nonstrange quark matter ($ud$QM) might be more stable~\cite{Holdom2018_PRL120-222001}. Similar to SQM, there may be stable $ud$QM nuggets~\cite{Xia2020_PRD101-103031, Wang2021_Galaxies9-70, Xia2022_PRD106-034016} and $ud$QM stars~\cite{Zhao2019_PRD100-043018, Zhang2020_PRD101-043003, Cao2022_PRD106-083007, Zhang2021_PRD103-063018, Yuan2022_PRD105-123004} existing in the Universe, while at $A\lesssim 300$ ordinary nuclei do not decay into $ud$QM nuggets due to finite size effects~\cite{Holdom2018_PRL120-222001, Xia2020_PRD101-103031, Wang2021_Galaxies9-70, Xia2022_PRD106-034016}.

For strangelets and $ud$QM nuggets, it was shown that the interface effects play important roles on their properties~\cite{Xia2019_AIP2127-020029, Xia2020_PRD101-103031, Lugones2016_EPJA52-53}, where the energy contribution can be treated with a surface tension $\sigma$. For example, in the framework of MIT bag model, it was found that small strangelets are destabilized significantly if $\sigma^{1/3} \approx B^{1/4}$ with $B$ being the bag constant~\cite{Farhi1984_PRD30-2379}, while the minimum baryon number for metastable strangelets $A_\mathrm{min} \propto \sigma^{3}$~\cite{Berger1987_PRC35-213, Berger1989_PRD40-2128}. If the surface tension $\sigma$ falls below a critical value $\sigma < \sigma_\mathrm{crit}$ with $\sigma_\mathrm{crit}$ typically on the order of a few MeV/$\mathrm{fm}^2$, then large strangelets become unstable and will fission into smaller and more stable ones~\cite{Alford2006_PRD73-114016}. Similarly, the surfaces of strange stars will fragment into crystalline crusts comprised of positively-charged strangelets embedded in a neutralizing background of electrons~\cite{Alford2008_PRC78-045802}. For $ud$QM nuggets, given the absence of $s$ quarks, the critical surface tension $\sigma_\mathrm{crit}$ is larger than that of strangelets~\cite{Alford2006_PRD73-114016, Jaikumar2006_PRL96-041101, Alford2008_PRC78-045802}, which may becomes even larger if there exist large symmetry energy in $ud$QM~\cite{Chu2014_APJ780-135, JEONG2016_NPA945-21, Chu2019_PRC99-035802, Wu2019_AIP2127_020032}. Under these conditions, it is easier for $ud$QM to reach the condition $\sigma < \sigma_\mathrm{crit}$, which indicates the possible existence of crusts in $ud$QM stars. Finally, it is worth mentioning that the precise value of $\sigma$ remains elusive but recent estimations suggest $\sigma \lesssim$ 30 MeV/$\mathrm{fm}^2$~\cite{Oertel2008_PRD77-074015, Palhares2010_PRD82-125018, Pinto2012PRC86-025203, Kroff2015_PRD91-025017, Ke2014_PRD89-074041, Mintz2013_PRD87-036004, Gao2016_PRD94-094030, Xia2018_PRD98-034031, You2024_PRD109-034003, Fraga2019_PRD99-014046, Lugones2017_PRD95-015804, Lugones2019_PRD99-035804}. For small strangelets and $ud$QM nuggets, beside the surface tension, the curvature term also play important roles~\cite{Madsen1993_PRL70-391, Madsen1993_PRD47-5156, Madsen1994_PRD50-3328, Lugones2021_PRC103-035813}, while $\sigma$ may increase substantially due to new terms stemming from vector interactions, ultimately hindering the formation of quark star crusts and compact dwarfs~\cite{Lugones2013_PRC88-045803, Lugones2021_PRC103-035813}. In this work, the curvature and surface contributions in strangelets and $ud$QM nuggets are considered self-consistently by solving the Dirac equations for quarks in the mean-field approximation (MFA), where an equivparticle model is adopted with both the linear confinement and leading-order perturbative
interactions~\cite{Peng1999_PRC61-015201, Chen2012_CPC36-947, Xia2014_PRD89-105027, Xia2018_PRD98-034031, Xia2019_AIP2127-020029, Xia2022_PRD106-034016, You2024_PRD109-034003}.

In violent astrophysical events such as strange star collisions, more than 0.03 $M_{\odot}$ of SQM may be ejected into space~\cite{Bauswein2009_PRL103-011101, Zhu2021_PRD104-083004, Zhou2022_PRD106-103030, Lin2023_PRD108-064007}. Then those ejected material may form strangelet dwarfs and strangelet planets if $\sigma < \sigma_\mathrm{crit}$, where the matter contents are similar to those in strange star crusts~\cite{Alford2012_JPG39-065201}. Similarly, if $ud$QM is more stable, there could be $ud$QM dwarfs and $ud$QM planets instead~\cite{Wang2021_Galaxies9-70, Xia2022_PRD106-034016, Liu2023_AAS68-1}. We refer to strangelet dwarfs and $ud$QM dwarfs as compact dwarfs, whose radii are typically smaller than traditional white dwarfs~\cite{Alford2012_JPG39-065201, Wang2021_Galaxies9-70, Xia2022_PRD106-034016, Liu2023_AAS68-1}. Additionally, a compact dwarf may be formed from dark matter if it was made of strangelets or $ud$QM nuggets~\cite{Zhitnitsky2003_JCAP2003-010, Banerjee2003_MNRAS340-284, Forbes2010_PRD82-083510}. There may be some observational hints for the possible existence of compact dwarfs. For example, the existence of possible ultra-low mass and small-radius white dwarfs with masses ranging from approximately 0.02 $M_{\odot}$ to 0.08 $M_{\odot}$ and radii spanning 4,270 km to 10,670 km~\cite{Kurban2022_PLB832-137204}. The observed underluminous Type Ia supernovae with progenitor masses seemingly far below the 1.4 $M_{\odot}$ threshold, thereby posing challenges to traditional white dwarf models predicting a Chandrasekhar mass limit of approximately 1.4 $M_{\odot}$~\cite{Flors2019_MNRAS491-2902}. Note that those white dwarfs may also be strange dwarfs consisting of a SQM core and a thick crust of normal matter~\cite{Glendenning1995_PRL74-3519, Geng2021_Innovation2-100152}, anisotropic magnetized white dwarfs~\cite{Deb2022_APJ926-66}, or white dwarfs composed primarily of heavy elements~\cite{Xia2023_FASS10-2296}.

Previous investigations on compact dwarfs assumes a small surface tension $\sigma < \sigma_\mathrm{crit}$~\cite{Alford2012_JPG39-065201, Wang2021_Galaxies9-70, Xia2022_PRD106-034016, Liu2023_AAS68-1}, so there exist stranglets and $ud$QM nuggets at certain sizes ($A\approx 1000$) that are more than others, in resemblance to finite nuclei with $^{56}$Fe being the most stable nucleus. Thus previous studies consider only the cases where the compact dwarfs are mostly made of those stable stranglets or $ud$QM nuggets. Nevertheless, such a condition does not need to be satisfied for compact dwarfs to exist stably. In fact, due to the presence of Coulomb barriers $E_\mathrm{C}$ among stranglets or $ud$QM nuggets, the fusion reactions does not take place for compact dwarfs with temperature $T\ll E_\mathrm{C}$~\cite{Xia2017_NPB916-669}. Then as long as  stranglets or $ud$QM nuggets do not decay into nuclei, compact dwarfs are stable if electron capture or $\beta$-decay reactions do not take place.
Similarly, if compact dwarfs are covered by a layer of normal matter, the fusion reaction between nuclei and stranglets/$ud$QM nuggets do not take place.

In this work, based on equivparticle model, we calculate the properties of $ud$QM nuggets and strangelets with different baryon ($A$) and charge ($Z$) numbers, where the corresponding mass charts are fixed. We then explore the possible existence of compact dwarfs comprised of $ud$QM nuggets or strangelets. It is found that compact dwarfs are mostly made of $ud$QM nuggets or strangelets near the $\beta$-stability line. The structures of compact (hybrid) dwarfs covered by normal matter ($^{16}\mathrm{O}$) are investigated as well. The stability of $ud$QM dwarfs, strangelet dwarfs, and hybrid dwarfs are then examined by solving the relativistic radial oscillation equations.

This paper is organized as follows. In Sec.~\ref{sec:the_nuggets}, we fix the density profiles and masses of strangelets and $ud$QM nuggets by solving the Dirac equations in the framework of the equivparticle model. In Sec.~\ref{sec:the_dwarfs}, based on the BPS (Baym-Pethick-Sutherland) model~\cite{Baym1971_APJ170-299}, we examine all possible candidates of $ud$QM nuggets and strangelets that could stably exist in compact dwarfs, where the equations of state (EOS) and the corresponding mass-radius relations  are fixed. The radial oscillation frequencies of fundamental and excited modes of compact dwarfs are discussed in detail in Sec.~\ref{sec:Oscillation}. Finally, we give our conclusion and outlook in Sec.~\ref{sec:con}.

\section{\label{sec:the_nuggets} light-quark nuggets}
To fix the properties of light-quark nuggets such as $ud$QM nuggets and strangelets, we adopt an equivparticle model~\cite{Peng1999_PRC61-015201, Chen2012_CPC36-947, Xia2014_PRD89-105027, Xia2018_PRD98-034031, Xia2019_AIP2127-020029, Xia2022_PRD106-034016, You2024_PRD109-034003}, where the strong interactions are considered with density-dependent quark masses and quarks are viewed as quasi-free particles. The Lagrangian density of the equivparticle model is given by
\begin{equation}
\mathcal{L} =  \sum_{i=u,d,s} \bar{\psi}_i \left[ i \gamma^\mu \partial_\mu - m_i(n_\mathrm{b}) - e q_i \gamma^\mu A_\mu \right]\psi_i
             - \frac{1}{4} A_{\mu\nu}A^{\mu\nu},  \label{eq:Lgrg_all}
\end{equation}
where $\psi_i$ is the Dirac spinor of quark flavor $i$, $m_i$ the mass,  $q_u=2/3$ and $q_d=q_s=-1/3$ the charge, and $A_\mu$ the photon field with the field tensor
\begin{equation}
A_{\mu\nu} = \partial_\mu A_\nu - \partial_\nu A_\mu.
\end{equation}
In the equivparticle model, by incorporating both linear confinement and leading-order perturbative interactions, the quark mass scaling is given by~\cite{Xia2014_PRD89-105027}
\begin{equation}
m_i(n_{\mathrm{b}})=m_{i0}+m_\mathrm{I}(n_{\mathrm b})= m_{i0}+Dn_{\mathrm b}^{-1/3} + Cn_{\mathrm b}^{1/3}, \label{Eq:D}
\end{equation}
where $m_{u0} = 2.2$ MeV, $m_{d0} = 4.7$ MeV, and $m_{s0} = 96.0$ MeV are the current quark masses~\cite{Patrignani2016_CPC40-100001}. The baryon number density is determined by $n_{\mathrm{b}} = \sum_{i=u,d,s}n_i/3$ with the quark number density $n_i = \langle\bar{\psi}_i\gamma^0\psi_i\rangle$. The confinement parameter $D$ is connected to the string tension, the chiral restoration density, and the sum of the vacuum chiral condensates~\cite{Peng1999_PRC61-015201}. Depending on the sign of $C$, the last term of Eq.~(\ref{Eq:D}) corresponds to the contribution of one-gluon-exchange interaction ($C<0$)~\cite{Chen2012_CPC36-947} or leading-order perturbative interaction ($C>0$)~\cite{Xia2014_PRD89-105027}. In this work, as an example, we adopt the parameter set $C=0.1$ and $\sqrt{D}=150$ MeV, which gives satisfactory predictions on light-quark nuggets and compact stars~\cite{Xia2022_PRD106-034016}. By employing the mean-field and no-sea approximations, the Dirac equations for quarks and the Klein-Gordon equation for photons are derived through a variational process. It is worth noting that electrons are ignored in this context due to their comparatively minor contributions in $ud$QM nuggets or strangelets with radii $R\lesssim40$ fm~\cite{Xia2017_JPCS861-012022}.

For spherically symmetric $ud$QM nuggets or strangelets, the Dirac spinor of quarks can be expanded as
\begin{equation}
 \psi_{n\kappa m}({\bm r}) =\frac{1}{r}
 \left(\begin{array}{c}
   iG_{n\kappa}(r) \\
    F_{n\kappa}(r) {\bm\sigma}\cdot{\hat{\bm r}} \\
 \end{array}\right) Y_{jm}^l(\theta,\phi)\:,
\label{EQ:RWF}
\end{equation}
where $G_{n\kappa}(r)/r$ and $F_{n\kappa}(r)/r$ are the radial wave functions for upper and lower components, while $Y_{jm}^l(\theta,\phi)$ is the spinor spherical harmonics. The quantum number $\kappa$ is defined by the angular momenta ($l,j$) as $\kappa = (-1)^{j+l+1/2}(j+1/2)$ with $j=l\pm1/2$.

By utilizing the mean-field approximation and inserting Eq.~(\ref{EQ:RWF}) into the Dirac equation, we can easily acquire the one-dimensional radial Dirac equation through the integration of the angular component, i.e.,
\begin{equation}
 \left(\begin{array}{cc}
  V_{iV} + V_{iS}                                                   & {\displaystyle -\frac{\mbox{d}}{\mbox{d}r} + \frac{\kappa}{r}}\\
  {\displaystyle \frac{\mbox{d}}{\mbox{d}r}+\frac{\kappa}{r}} & V_{iV} - V_{iS}                       \\
 \end{array}\right)
 \left(\begin{array}{c}
  G_{in\kappa} \\
  F_{in\kappa} \\
 \end{array}\right)
 = \varepsilon_{in\kappa}
 \left(\begin{array}{c}
  G_{in\kappa} \\
  F_{in\kappa} \\
 \end{array}\right) \:,
\label{Eq:RDirac}
\end{equation}
where $\varepsilon_{in\kappa}$ is the single particle energy. The mean-field scalar and vector potentials of quarks can be obtained as
\begin{eqnarray}
 V_{iS} &=& m_{i0} +
 m_\mathrm{I}(n_\mathrm{b}), \label{Eq:Vs}\\
 V_{iV} &=& \frac{1}{3}\frac{\mbox{d} m_\mathrm{I}}{\mbox{d} n_\mathrm{b}}\sum_{i=u,d,s}  n_i^\mathrm{s} + e q_i A_0, \label{Eq:Vv} \label{Eq:Vv}
\end{eqnarray}
where there exist common terms for the scalar and vector potentials $V_{S} = m_\mathrm{I}(n_\mathrm{b})$ and $V_{V} = \frac{1}{3}\frac{\mbox{d} m_\mathrm{I}}{\mbox{d} n_\mathrm{b}}\sum_{i=u,d,s}  n_i^\mathrm{s}$.
In addition, it should be noted that the scalar potential in Eq.~(\ref{Eq:Vs}) now incorporates the current masses of quarks. The vector potential arises as a result of the density-dependence of quark masses and is crucial for ensuring thermodynamic self-consistency~\cite{Xia2018_PRD98-034031,Xia2014_PRD89-105027}. The Klein-Gordon equation for photons is given by
\begin{equation}
- \nabla^2 A_0 = e n_\mathrm{ch}, \label{Eq:K-G}
\end{equation}
where $n_\mathrm{ch}=\sum_iq_in_i$ is the charge density. At given scalar and vector potentials, the single particle energies for quarks are obtained by solving the Dirac equation~(\ref{Eq:RDirac}), where for zero temperature cases quarks occupy the lowest energy levels according to Fermi-Dirac statistics. The corresponding scalar and vector densities of quark flavor $i$ are then fixed by
\begin{eqnarray}
 n_i^\mathrm{s}(r) &=& \sum_{k=1}^{N_i} \frac{1}{4\pi r^2}
 \left[|G_{ik}(r)|^2-|F_{ki}(r)|^2\right], \label{Eq:Densitys} \\
 n_i(r) &=& \sum_{k=1}^{N_i} \frac{1}{4\pi r^2}
 \left[|G_{ik}(r)|^2+|F_{ik}(r)|^2\right], \label{Eq:Densityv}
\end{eqnarray}
where $k$ represents the quantum state $(n,\kappa,m)$ of quarks. The total quark number $N_i$ is fixed by integrating the density $n_i(r)$ in coordinate space, i.e.,
\begin{equation}
N_i=\int 4\pi r^2 n_i(r)dr.
\end{equation}

At fixed baryon number $A$, we solve iteratively the Dirac Eq.~(\ref{EQ:RWF}), mean-field Eqs.~(\ref{Eq:Vs}) and (\ref{Eq:Vv}), Klein-Gordon Eq.~(\ref{Eq:K-G}) and density Eqs.~(\ref{Eq:Densitys}) and (\ref{Eq:Densityv}) inside a box with grid width less than 0.005 fm. Once convergency is reached, the total mass of a strangelet or $ud$QM nugget can be obtained with
\begin{equation}
M = \sum_i\sum_{k=1}^{N_i} \varepsilon_{ik}  - \int 2\pi r^2 (6n_\mathrm{b} V_V + e n_\mathrm{ch}  A_0) \mbox{d}r. \label{Eq:M}
\end{equation}

\begin{figure}
\includegraphics[width=\linewidth]{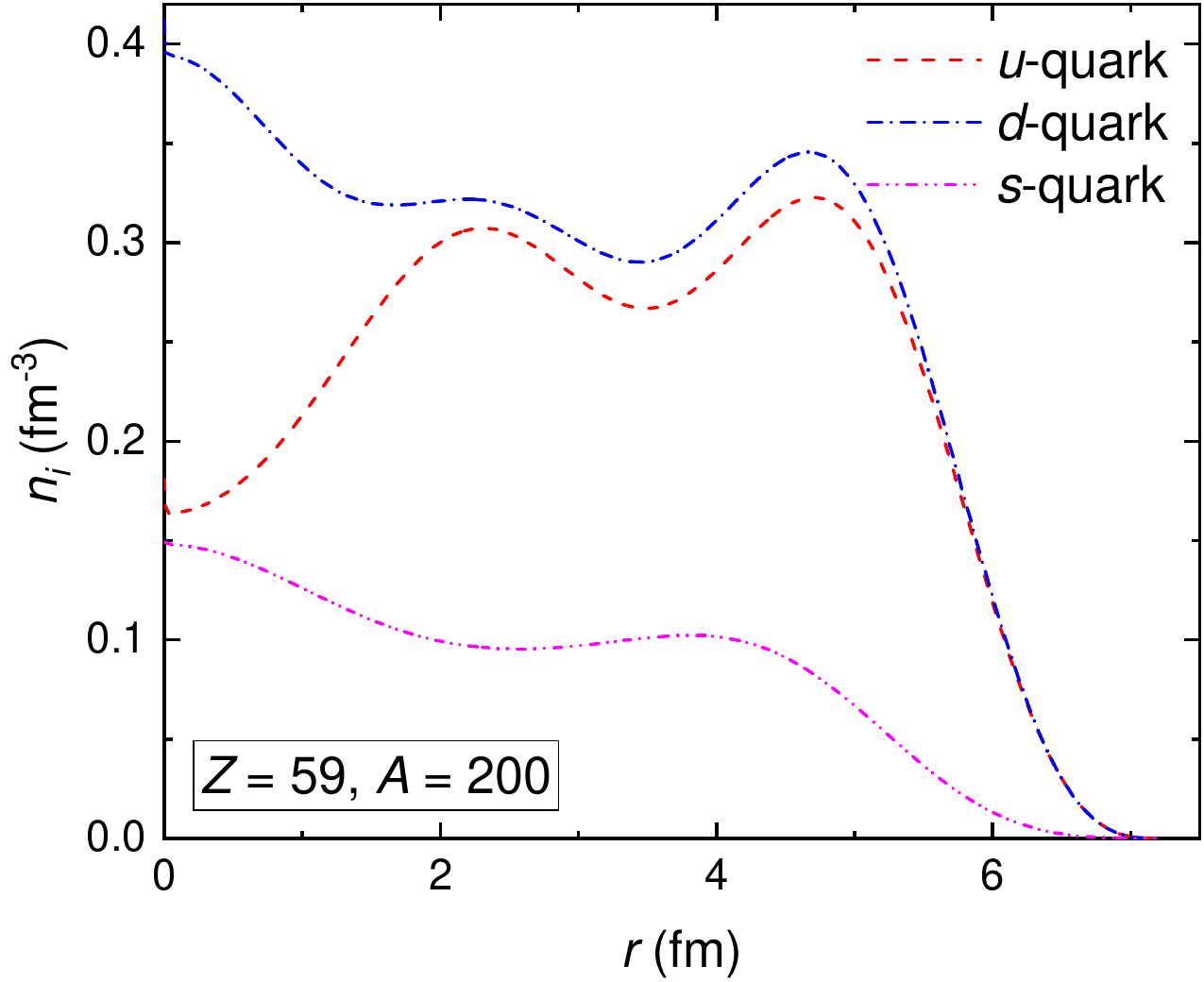}
\caption{\label{Fig:Dens} Density distributions of $u, d$, and $s$ quarks for a $\beta$-stable strangelet at $Z=59$ and $A=200$.}
\end{figure}

In Fig.~\ref{Fig:Dens}, as an example, we present the density distributions of $u$, $d$, and $s$ quarks for a $\beta$-stable strangelet at $Z=59$ and $A=200$. As quark wave functions contain multiple nodes, the density profiles derived from Eq.~(\ref{Eq:Densityv}) exhibit fluctuations as functions of the radial coordinate $r$. The density on the surface drops slowly to zero since we consider linear confinement with density dependent quark masses. We believe such a scenario is more reasonable~\cite{Xia2018_PRD98-034031}. The density of $d$ quarks is generally larger than $u$ and $s$ quarks, while the density of $s$ quarks is smallest. This is mainly attributed to the requirement of $\beta$-stability condition where the composition of $u, d$, and $s$ quarks are optimized to to minimize the total mass $M$ in Eq.~(\ref{Eq:M}) at a fixed $A$.

\begin{figure*}[htbp]
\begin{minipage}[t]{0.45\linewidth}
\centering
\includegraphics[width=\textwidth]{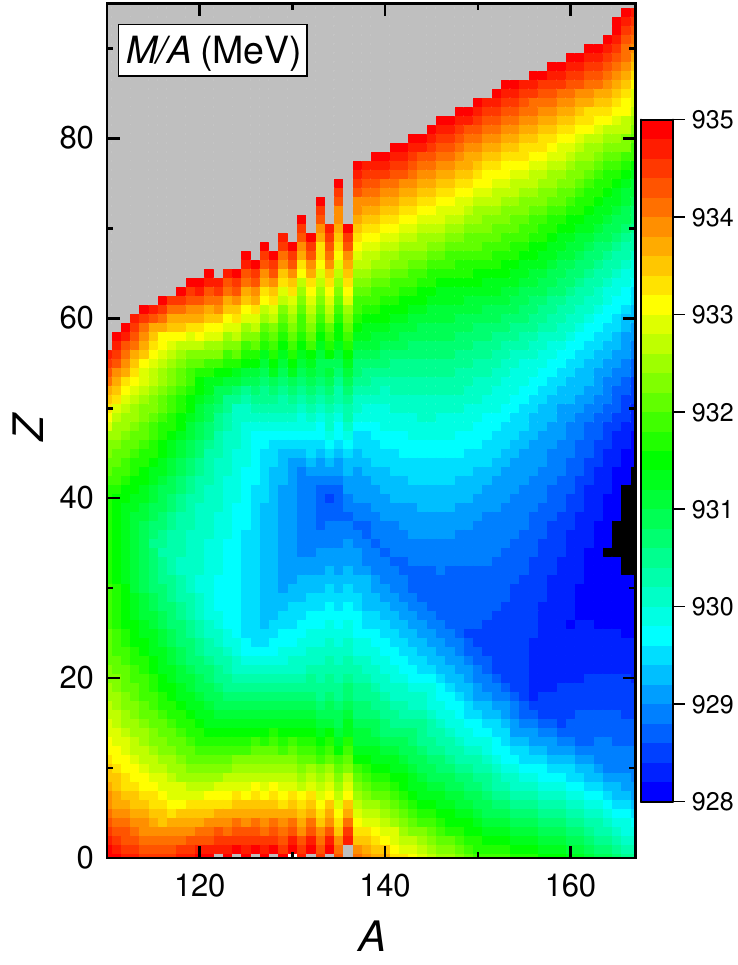}
\end{minipage}%
\hfill
\begin{minipage}[t]{0.53\linewidth}
\centering
\includegraphics[width=\textwidth]{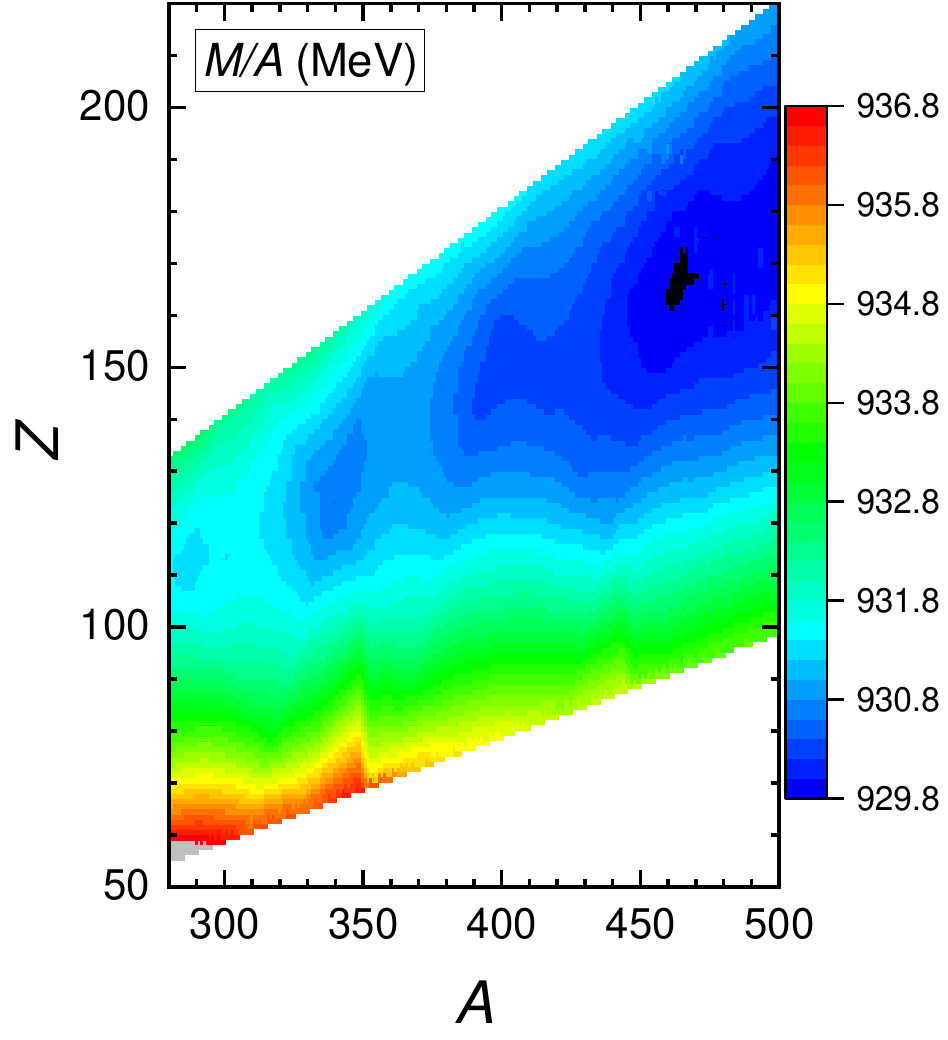}
\end{minipage}
\caption{\label{Fig:QM-chart} Energy per baryon of strangelets (left) and $ud$QM nuggets (right) at various baryon numbers $A$ and charge numbers $Z$, where the parameter set $C$ = 0.1 and $\sqrt{D} = 150$ MeV is employed.}
\end{figure*}

We then carry out systematic calculations on the properties of strangelets and $ud$QM nuggets at various $A$ and $Z$. In the left panel of Fig.~\ref{Fig:QM-chart}, the mass chart of strangelets with $Z=0$-100 and $A=110$-167 are presented, where the parameter set $C=0.1$ and $\sqrt{D}=150$ MeV is adopted. Note that the number of $s$ quarks $N_s$ is fixed by minimizing the mass at fixed $A$ and $Z$, i.e., the chemical equilibrium between $d$ and $s$ quarks is attained. At a fixed baryon number $A$, there exists a charge number $Z$ with a minimum energy per baryon, corresponding to the $\beta$-stable strangelet. It is evident that $\beta$-stable strangelets carry less charge than nuclei, which is mainly due to the emergence of $s$ quarks. The obtained strangelets becomes more stable if we increase $A$, which becomes absolutely stable at $A\geq 120$ with the energy per baryon fulfilling $M/A<930$ MeV. Note that throughout the mass chart of strangelets with various $A$ and $Z$, we observe strangelets at certain areas that are more stable than others. This is mainly attributed to shell effects, where various magic numbers with large shell gaps in the single-quark-energy levels were identified for $u$, $d$, $s$ quarks~\cite{Xia2019_AIP2127-020029, You2024_PRD109-034003}. Finally, we should mention that exist slight numerical uncertainties ($\Delta M/A \lesssim 1$ MeV) for strangelets far from $\beta$-stability line at $A\approx 120$-140. This is mainly attributed to the iteration method used here, where the converged results correspond to local energy minima instead of the actual  ground state. Nevertheless, this has little impact for our current study on strangelet dwarfs, where only the strangelets located around $\beta$-stability line emerge.

By taking $N_s=0$, the properties of $ud$QM nuggets can then be obtained. In the right panel of Fig.~\ref{Fig:QM-chart} we present the mass chart for $ud$QM nuggets, where the energy per baryon at baryon numbers $A=280$-500 and various charge numbers $Z$ around the $\beta$-stability line are indicated. Note that a symmetry energy correction term $\Delta M_\mathrm{sym}$ is added to Eq.~(\ref{Eq:M}) for $ud$QM nuggets with
\begin{equation}
\Delta M_\mathrm{sym}= \frac{a_\mathrm{sym}}{A} \left( \frac{A}{2} - Z \right)^2,
\end{equation}
where the parameter $a_\mathrm{sym}=65$ MeV is adopted so that $ud$QM nuggets is unstable compared to finite nuclei. The variation of energy per baryon for $ud$QM nuggets generally resembles the trend of strangelets except that $\beta$-stable $ud$QM nuggets carry more charge than strangelets, which is due to the absence of $s$ quarks. As baryon number $A$ increases, the energy per baryon of $ud$QM nuggets generally decreases and eventually becomes smaller than 930 MeV at $A\gtrsim 450$, i.e., the absolute stability of $ud$QM. A stability island at $A=340$ and $Z=122$ is observed for $ud$QM nuggets, which is attributed to the shell effects.  Nevertheless, the obtained magic numbers with large shell gaps may be altered if we adopt different parameter sets~\cite{Xia2019_AIP2127-020029}.

\begin{figure}
\includegraphics[width=\linewidth]{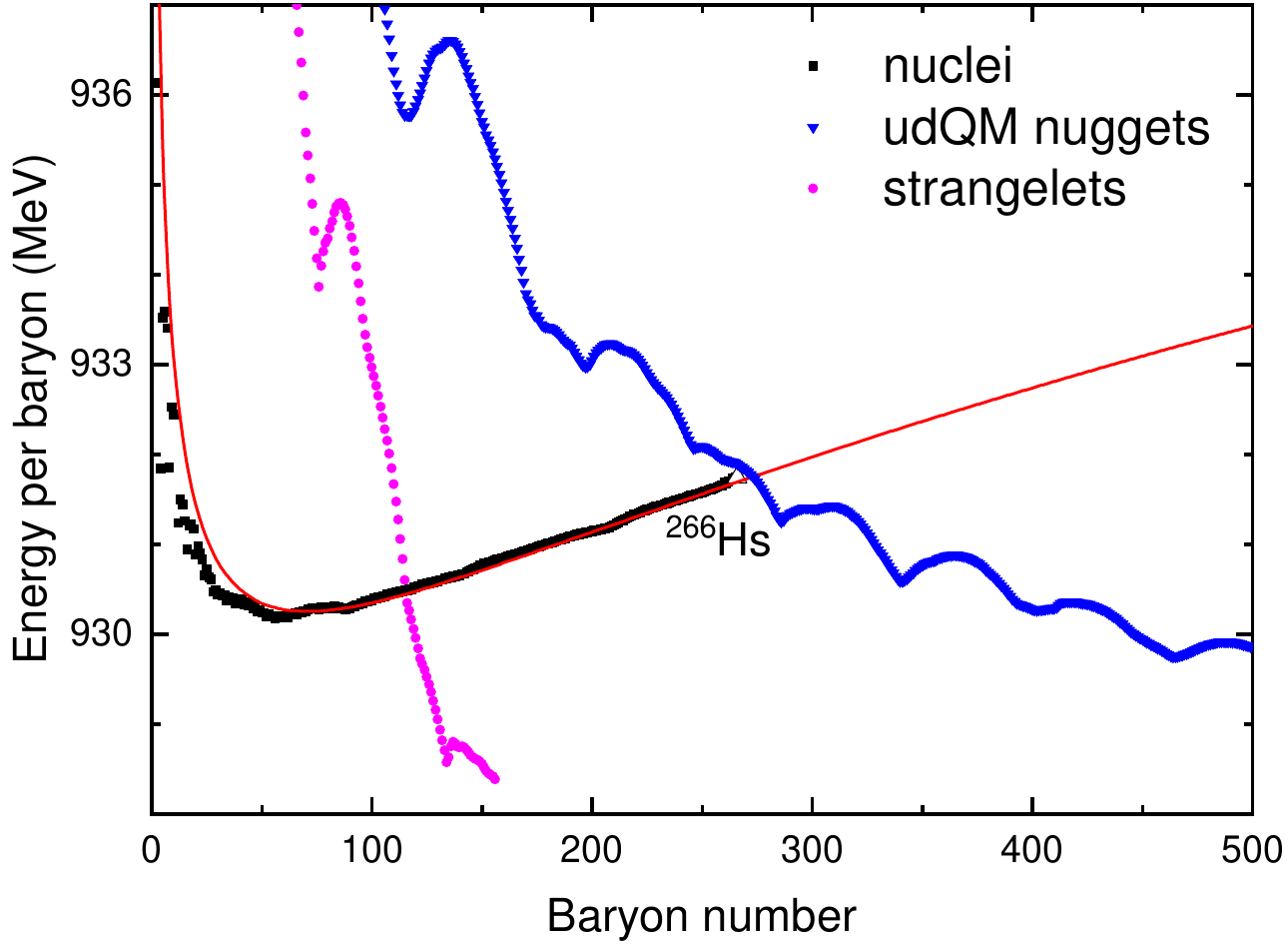}
\caption{\label{Fig:beta-stable} Energy per baryon of $\beta$-stable $ud$QM nuggets, strangelets and nuclei as functions of baryon number $A$. The red solid curve indicates the energy per baryon of $\beta$-stable nuclei obtained by the liquid-drop formula in Eq.~(\ref{Eq:LD}).}
\end{figure}

Based on the mass tables indicated in Fig.~\ref{Fig:QM-chart} as well as the AME 2020 atomic mass evaluation~\cite{Huang2021_CPC45-030002,Wang2021_CPC45-030003}, we can then fix the charge numbers $Z$ and masses of $\beta$-stable $ud$QM nuggets, strangelets and nuclei by searching for the energy minima, where in Fig.~\ref{Fig:beta-stable} the corresponding energy per baryon are presented as functions of baryon number $A$. The red solid curve in Fig.~\ref{Fig:beta-stable} shows the energy per baryon of nuclei determined by a liquid-drop formula, i.e.,
\begin{equation}
  \varepsilon_\mathrm{LD} =   \varepsilon_\mathrm{NM} + \frac{\sqrt [3]{36 \pi n_0}\alpha Z^2}{5 A^{4/3}}  + \sigma\left(\frac{A n_0^2}{36\pi}\right)^{-1/3}, \label{Eq:LD}
\end{equation}
where on the right hand side are respectively the energy contributions of bulk nuclear matter, Coulomb interaction, and surface corrections. A more detail explanation of the formula and the exact values of the coefficients can be found in Ref.~\cite{Liu2023_AAS68-1}.

Compared to atomic nuclei, as indicated in Fig.~\ref{Fig:beta-stable}, $ud$QM nuggets and strangelets exhibit larger shell corrections, leading to more significant fluctuations in the energy per baryon as we vary baryon number $A$. The stability of strangelets or $ud$QM nuggets can also be identified easily. It is found that the energy per baryon for strangelets become smaller than that of finite nuclei at $A \geq 116$ and quickly drops below 930 MeV at $A \geq 120$, i.e., more stable than $^{56}$Fe. Similarly, according to the liquid-drop formula in Eq.~(\ref{Eq:LD}), $ud$QM nuggets become more stable than nuclei at $A \geq 274$, while at $A \geq 447$ the energy per baryon drops below 930 MeV. Since finite nuclei do not decay into $ud$QM nuggets on Earth, the obtained masses of $ud$QM nuggets are consistent with the current experimental constraints, where the heaviest $\beta$-stable nucleus $^{266}\mathrm{Hs}$ is more stable than the corresponding $ud$QM nugget. For $ud$QM nuggets with $274\leq A \leq 446$, there are possibilities that they may decay into nuclei. However, since the surface tension of the $ud$QM nuggets is much larger than that of finite nuclei, the preformation probability of nuclear clusters in $ud$QM nuggets maybe insignificant, making those objects meta-stable. Note that when $A$ falls below 10, the center-of-mass correction and one-gluon-exchange interactions become important and reduce the masses of strangelets or $ud$QM nuggets significantly~\cite{Aerts1978_PRD17-260,You2022_NPR39-302}, which will not be considered in this work.

\section{\label{sec:the_dwarfs} Compact dwarfs}
With the properties of strangelets and $ud$QM nuggets fixed in Sec.~\ref{sec:the_nuggets}, we then investigate the matter contents and structures of compact dwarfs made of those objects. The cold compact dwarf matter is made of strangelets or $ud$QM nuggets embedded in a homogeneous electron gas, forming a body-centered cubic crystal lattice. Applying the BPS model~\cite{Baym1971_APJ170-299}, the energy density of compact dwarf matter can be obtained with
\begin{equation}
E = \frac{M n_e}{Z} + \left(1 + \frac{\alpha}{2\pi}\right)E_e - K_M \alpha \left(\frac{4 \pi n_e Z^2}{3}\right)^{1/3} \sigma(Z),
\label{Eq:E-D}
\end{equation}
where $n_e$ is the average electron number density and $\alpha = 1/137.03599911$ is the fine structure constant. The first term in the formula is the energy density of strangelets or $ud$QM nuggets fixed by Eq.~(\ref{Eq:M}) at a given set of charge ($Z$) and baryon ($A$) numbers. The baryon number density is given by $n_\mathrm{b} = n_e/f_Z$ with the charge-to-mass ratio $f_Z = Z/A$. Then the Wigner-Seitz (WS) cell radius is obtained with $R_W = (3A/4 \pi n_\mathrm{b})^{1/3}$. The second term in Eq.~(\ref{Eq:E-D}) is the energy density of electrons including exchange effect, where $E_e$ is the energy density of free electrons gas, i.e.,
\begin{equation}
E_e = \frac{m_e^4}{8\pi^2} \left[x_e(2x_e^2+1)\sqrt{x_e^2+1}-\mathrm{arcsh}(x_e) \right]. \label{Eq:e-E-D}
\end{equation}
Here $x_e \equiv (3\pi^2 n_e)^{1/3}/m_e$ with $m_e = 0.51099895$ MeV  being the electron mass. The third term in Eq.~(\ref{Eq:E-D}) accounts for the lattice energy density with electron polarization correction. In the context of a body-centered cubic lattice composed of strangelets or $ud$QM nuggets, we take the Madelung constant $K_M = -0.895929255682$~\cite{Baiko2001_PRE64-057402} and the dimensionless function~\cite{Potekhin2000_PRE62-8554, Chamel2020_PRC101-032801}
\begin{equation}
\sigma(Z) = 1 + \frac{12^{4/3}Z^{2/3}\alpha}{35\pi^{1/3}} \left(1-\frac{1.1866}{Z^{0.267}}+\frac{0.27}{Z}\right)
\label{Eq:E-Polar}
\end{equation}
In order for strangelets or $ud$QM nuggets to stably exist in compact dwarfs, they must be stable against $\beta$-decay and electron capture reactions~\cite{Xia2023_FASS10-2296}, i.e.,
\begin{eqnarray}
 && _Z^AX + e^- \rightarrow _{Z-1}^AX + \nu_e, \label{Eq:E-C} \\
 && _Z^AX \rightarrow _{Z+1}^AX +e^- +\overline{\nu}_e. \label{Eq:beta-decay}
\end{eqnarray}
When either one of the reactions occurs, the electron number density varies to $n_e(Z\pm1)/Z$ with the charge number of quark nuggets from $Z$ to $Z\pm1$. This indicates a stability criterion for the energy density, i.e.,
\begin{equation}
E(Z\pm1,A,n_b)-E(Z,A,n_b)>0. \label{Eq:E-D-Condition}
\end{equation}
For any sets of $Z$ and $A$ fulfilling this condition, $ud$QM nuggets or strangelets can exist stably in compact dwarfs. Then the energy density of compact dwarf matter can be determined by Eq.~(\ref{Eq:E-D}). In practice, we also search for the $ud$QM nuggets or strangelets that minimizes the energy density at a fixed baryon number density $n_\mathrm{b}$ with $n_\mathrm{b} \approx 10^{-15}\sim 1$ $\mathrm{fm}^{-3}$. Note that at sufficiently small densities, our hypothesis of fully ionized $ud$QM nuggets or strangelets no longer holds, where some of the electrons are trapped in the inner orbits and they are partially ionized. Meanwhile, at exceedingly large densities, compact dwarf matter will undergo a first-order phase transition to form uniform quark star matter. For simplicity, we do not consider those effects in our current study. Finally, the pressure is obtained according to the basic thermodynamic relations, i.e.,
\begin{equation}
P = n_{\mathrm{b}} \frac{\mathrm{d}E}{\mathrm{d}n_{\mathrm{b}}}-E. \label{Eq:P-nb}
\end{equation}

\begin{figure}
\includegraphics[width=\linewidth]{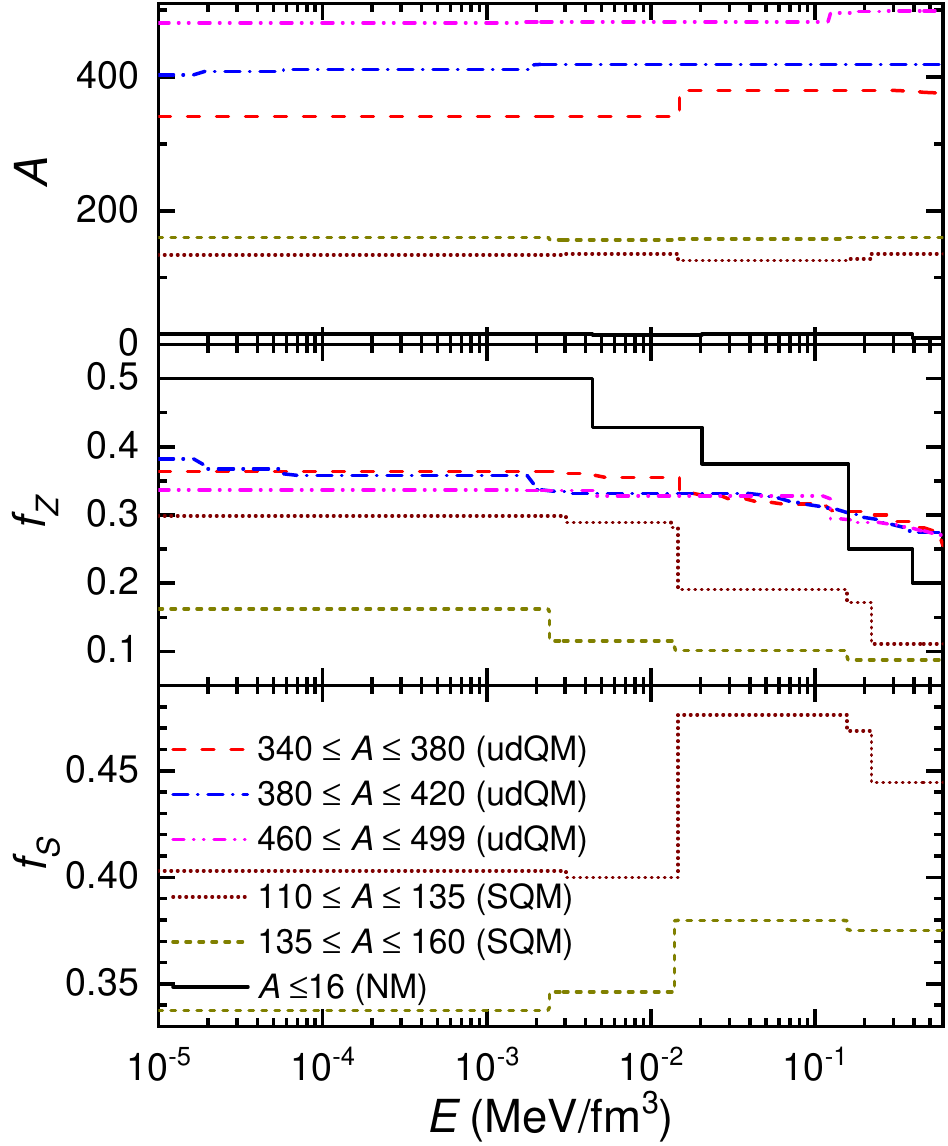}
\caption{\label{Fig:fz-fs} Baryon number $A$, charge-to-mass ratio $f_Z=Z/A$ and strangeness per baryon $f_S=N_s/A$ of strangelets and $ud$QM nuggets in compact dwarfs under different considerations, which are fixed employing the mass chart indicated in Fig.~\ref{Fig:QM-chart}. For comparison, the nuclear species with $A\leq 16$ in normal white dwarfs are presented as well~\cite{Xia2023_FASS10-2296}.}
\end{figure}

In Fig.~\ref{Fig:fz-fs}, we present the baryon numbers, charge-to-mass ratios, and strangeness per baryon for $ud$QM nuggets and strangelets that minimize the energy density of compact dwarf matter at fixed $n_\mathrm{b}$, assuming their baryon numbers locate within $340 \leq A \leq 380$, $380 \leq A \leq 420$, $460 \leq A \leq 499$ for $ud$QM dwarfs and $110 \leq A \leq 135$, $135 \leq A \leq 160$ for strangelet dwarfs. It is found that these $ud$QM nuggets and strangelets generally locate in the vicinity of $\beta$-stability line in Fig.~\ref{Fig:QM-chart}. At small densities with $E \lesssim 10^{-5}$ MeV$/\mathrm{fm}^{3}$, the species of strangelets, $ud$QM nuggets, and nuclei remain unchanged. In general, the $ud$QM nuggets and strangelets in compact dwarfs exhibit lower charge-to-mass ratios in comparison with nuclei in normal white dwarfs. This is mainly because the charge-to-mass ratios of $\beta$-stable $ud$QM nuggets and strangelets are reduced at larger $A$, while the presence of $s$ quarks in strangelets substantially reduce the mass and charge. Note that $f_Z$ does not always decrease as we increase $A$. For example, the charge-to-mass ratios of $ud$QM nuggets in the range of $340 \leq A \leq 380$ become larger at $E \lesssim 5\times10^{-5}$ MeV$/\mathrm{fm}^{3}$  if we consider $ud$QM nuggets within $380 \leq A \leq 420$, which is due to the stability island located at $A=340$ and $Z=122$ in the right panel of Fig.~\ref{Fig:QM-chart}. At larger densities, the species of $ud$QM and strangelets begins to change, which  typically occurs in the central regions of $ud$QM dwarfs and strangelet dwarfs. For strangelet dwarfs, the strangeness per baryon $f_S$ increases if we take larger $A$, while the corresponding strangelets become more stable.

\begin{figure}
\includegraphics[width=\linewidth]{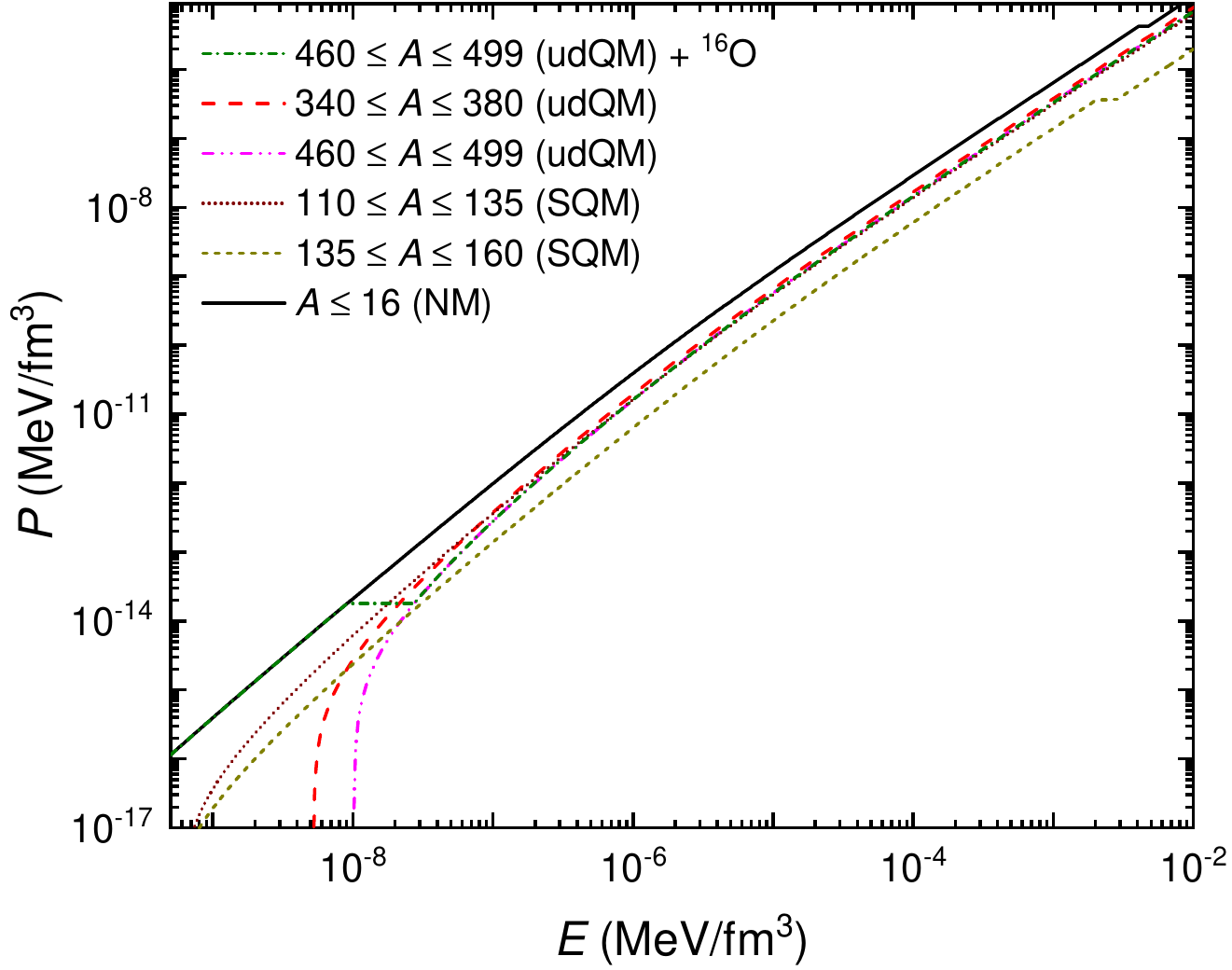}
\caption{\label{Fig:EOS} Pressure of compact dwarf matter as functions of energy density, where the corresponding species of $ud$QM nuggets and strangelets are indicated in Fig.~\ref{Fig:fz-fs}.}
\end{figure}

With the species of $ud$QM nuggets and strangelets fixed in Fig.~\ref{Fig:fz-fs}, the pressure of compact dwarf matter can be estimated using Eq.~(\ref{Eq:P-nb}), where the corresponding EOSs under various constraints are presented in Fig.~\ref{Fig:EOS}. Generally speaking, the pressure of compact dwarf matter is significantly smaller than that of white dwarf matter according to BPS model. At energy density $E \gtrsim 10^{-3}\ \mathrm{MeV}/\mathrm{fm}^{3}$, the EOSs of $ud$QM dwarf matter tend to coincide with each other due to similar charge-to-mass ratios, except for the minor variations stemming from the abrupt change in $ud$QM nugget species, leading to subtle first-order phase transitions. The EOSs of strangelet dwarf matter are also dominated by the charge-to-mass ratios, where small $f_Z$ leads to a smaller pressures. In fact, a reduction of the charge-to-mass ratio leads to a diminished electron number density with $n_e = n_\mathrm{b}f_Z$, thereby reducing the lattice and electron energy densities and ultimately the pressure is reduced. In addition to compact dwarf matter made entirely of $ud$QM nuggets or strangelets, we consider also the case where compact dwarfs are covered by normal matter ($^{16}$O) assuming $P = 10^{-14}\ \mathrm{MeV}/\mathrm{fm}^{3}$ at the interface.

\begin{figure*}[htbp]
\begin{minipage}[t]{0.48\linewidth}
\centering
\includegraphics[width=\textwidth]{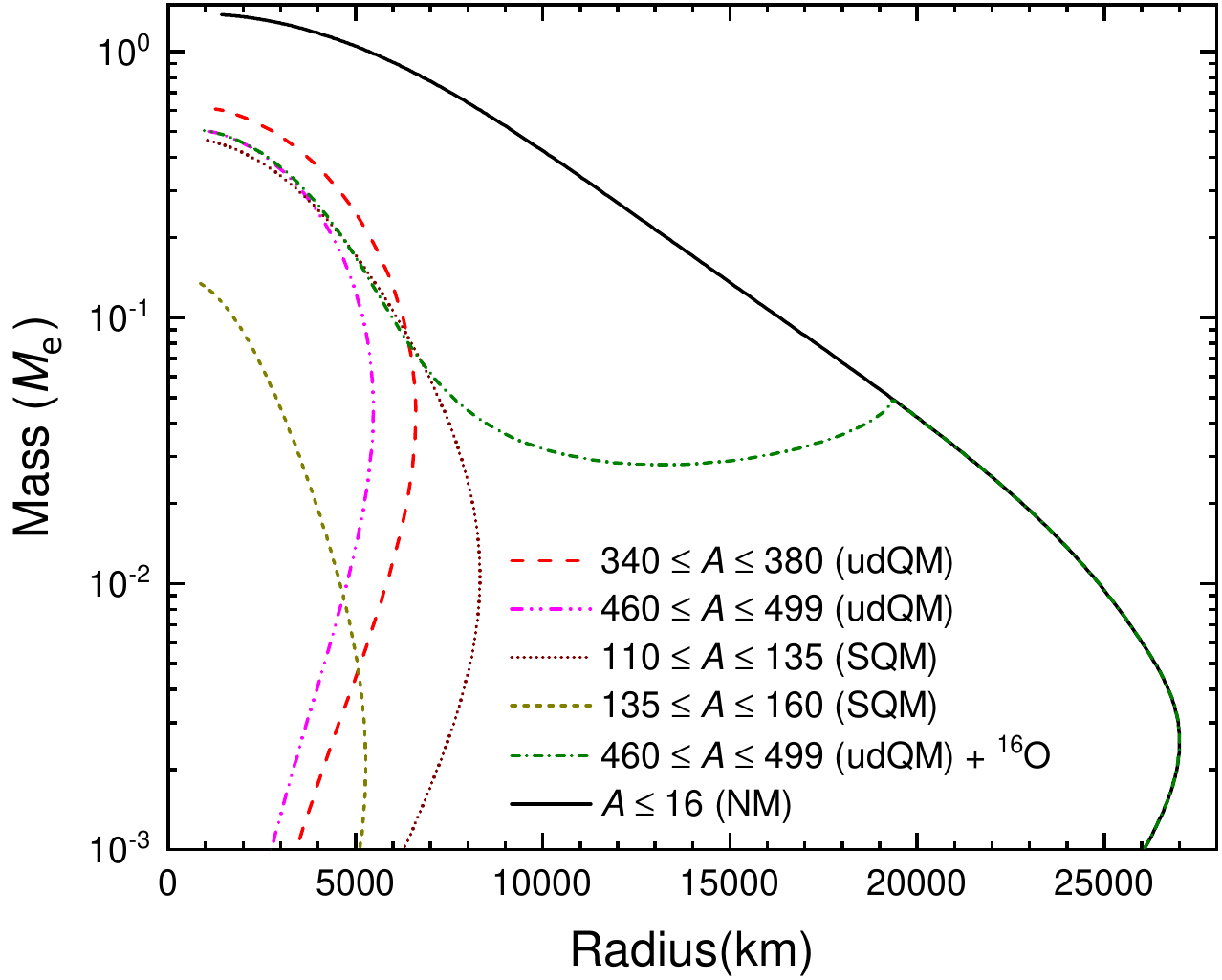}
\end{minipage}%
\hfill
\begin{minipage}[t]{0.48\linewidth}
\centering
\includegraphics[width=\textwidth]{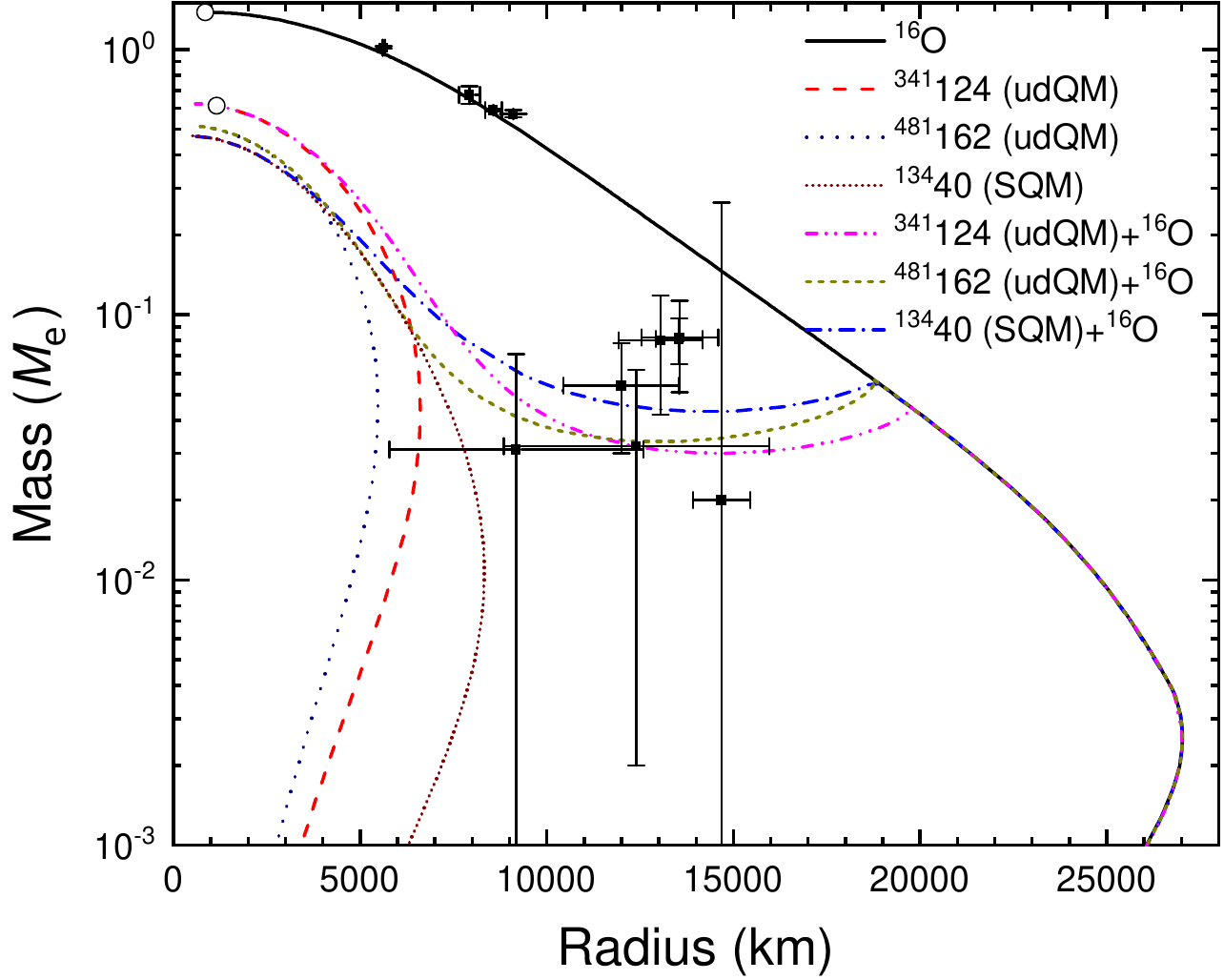}
\end{minipage}
\caption{\label{Fig:MR} Mass-radius relations of compact dwarfs obtained with the species indicated in Fig.~\ref{Fig:fz-fs} (left) and one particular $ud$QM nugget or strangelet (right), while the $ud$QM and strangelet dwarfs with surfaces covered by $^{16}\mathrm{O}$ are presented as well. For fixed species in the right pannel, the open circles indicate the critical masses where more massive stars become unstable against electron capture reactions. The four squares represent typical white dwarfs with masses exceeding 0.5 $\mathrm{M}_\odot$~\cite{Bond_2017_Astrophysical}, while the seven squares positioned below depict white dwarfs with unusually small masses and radii~\cite{Kurban2022_PLB832-137204}.}
\end{figure*}

Although the general relativistic effect of compact dwarfs is minute and can be adequately accurate using Newtonian gravity, we opt for a more comprehensive application of general relativity for our calculations. To fix the structures of compact dwarfs, we first present the following line element with static and spherically symmetric metric, i.e.,
\begin{equation}
\mathrm{d}s^2 = -e^{2\Phi}\mathrm{d}t^2 + e^{2\Lambda}\mathrm{d}r^2 + r^2(\mathrm{d}\theta^2 + \sin^2\theta\mathrm{d}\phi^2), \label{Eq:L-M}
\end{equation}
where $\Phi$ and $\Lambda$ are the metric functions of $r$. Assuming compact dwarf matter can be described by a perfect fluid with the energy-momentum tensor
\begin{equation}
T_{\mu\nu} = (E+P)u_{\mu}u_{\nu}+ P g_{\mu\nu}. \label{Eq:E-M-T}
\end{equation}
A mass function is defined as $m(r) = r(1-e^{-2\Lambda})/2$, which satisfies
\begin{equation}
\frac{\mathrm{d}m}{\mathrm{d}r} = 4 \pi r^2 E. \label{Eq:M-rho}
\end{equation}
Then the Tolman-Oppenheimer-Volkoff (TOV) equations that determine the pressure $P(r)$ and metric function $\Phi(r)$ are expressed as
\begin{eqnarray}
  \frac{\mathrm{d}P}{\mathrm{d}r} &=& -(E+P)\frac{\mathrm{d}\Phi}{\mathrm{d}r}, \label{Eq:P-r} \\
  \frac{\mathrm{d}\Phi}{\mathrm{d}r} &=& \frac{G m+4\pi r^3 GP}{r(r-2Gm)}. \label{Eq:phi-r}
\end{eqnarray}
Here $G = 6.707\times 10^{-45}\mathrm{MeV}^{-2}$ is the gravitational constant. By integrating Eq.~(\ref{Eq:M-rho}-\ref{Eq:phi-r}), we can then fix the mass $M=m(R)$, radius $R$, and metric elements ($e^{2\Phi}$, $e^{2\Lambda}$) of a compact dwarf.

In Fig.~\ref{Fig:MR} we present the mass-radius relations of $ud$QM dwarfs, strangelet dwarfs and hybrid dwarfs, taking into account various constraints on their matter compositions as illustrated in Fig.~\ref{Fig:fz-fs} as well as with one particular $ud$QM nugget or strangelet. The solid black curves in Fig.~\ref{Fig:MR} outlines the mass-radius relations of normal white dwarfs ($^{16}\mathrm{O}$), which are consistent with the observational masses and radii of the four white dwarfs with masses exceeding 0.5 $\mathrm{M}_\odot$. Compared with normal white dwarfs, $ud$QM dwarfs and strangelet dwarfs are significantly more compact with much smaller masses and radii. The inclusion of $ud$QM nuggets and strangelets with smaller $A$ leads to larger masses and radii for compact dwarfs~\cite{Liu2023_AAS68-1}, but still smaller than those of normal white dwarfs. Note that there exist minimum baryon numbers $A_\mathrm{min}= 274$ and 116 for $ud$QM nuggets and strangelets, which are unstable compared with nuclei of same $A$. At $A_\mathrm{min}\leq A \leq A_\mathrm{stb}$ with $A_\mathrm{stb}=273$ and 119 for $ud$QM nuggets and strangelets, they are metastable with respect to $^{56}$Fe. The quark nuggets become absolutely stable against decaying into nuclei only at $A > A_\mathrm{stb}$, while there are still possibilities for $\beta$-decay or electron capture reactions to take place in compact dwarfs. In general, at $A\lesssim 1000$, larger $ud$QM nuggets or strangelets are more stable, so that they are unlikely to fission nor decaying into atomic nuclei inside compact dwarfs. Meanwhile, the charge-to-mass ratios of quark nuggets tend to decrease at larger $A$, which leads to a reduction of masses and radii for compact dwarfs. If we consider compact dwarfs cover by normal white dwarf matter ($^{16}$O), the mass-radius relations of $ud$QM dwarfs and white dwarfs are then connected by such hybrid dwarfs. These hybrid dwarfs could be formed via accretion, while the likelihood of converting nuclei into $ud$QM nuggets or strangelets is exceedingly low due to the presence of Coulomb barriers. It is found that the hybrid dwarfs can account for the seven ultra-low-mass and small-radius white dwarfs~\cite{Kurban2022_PLB832-137204} if we assume the pressure at the interface of compact dwarf matter and normal matter is $P = 10^{-14} \ \mathrm{MeV}/\mathrm{fm}^3$.

\section{\label{sec:Oscillation} Radial oscillations}
By introducing a displacement $\delta r(r, t)$ to a fluid element in compact dwarfs, we can then investigate the radial oscillations, where the displacement oscillating at a frequency $\omega$ reads
\begin{equation}
\delta r(r,t) = X(r)e^{i\omega t}. \label{Eq:DELTA-r}
\end{equation}
Rewrite the displacement function as $\zeta = r^2 e^{-\Phi}X$ and adopt the perturbation equations in Ref.~\cite{Kokkotas2001_AA366-565}, the master equation of radial oscillations can be determined by
\begin{equation}
\frac{\mathrm{d}}{\mathrm{d}r}\left(\mathcal{P}\frac{\mathrm{d}\zeta}{\mathrm{d}r}\right) +(\mathcal{Q}+\omega^2\mathcal{W})\zeta = 0
\label{Eq:R-O-E}
\end{equation}
with
\begin{eqnarray}
  r^2\mathcal{P} &=& \Gamma P e^{\Lambda+3\Phi}, \label{Eq£ºP-fun} \\
  r^2\mathcal{Q} &=& e^{\Lambda+3\Phi}(E+P)\left[(\Phi^{\prime})^2 + 4\frac{\Phi^{\prime}}{r}-8\pi e^{2\Lambda}P\right], \label{Eq:Q-fun}\\
  r^2\mathcal{W} &=& (E+P)e^{3\Lambda+\Phi}. \label{Eq:W-fun}
\end{eqnarray}
Here the prime represents taking derivative with respect to $r$ and $\Gamma = \frac{E+P}{P} \frac{\mathrm{d}P}{\mathrm{d}E}$ is the adiabatic index of compact dwarf matter. By assigning the variable $\eta = \mathcal{P}$, one arrives at the following set of coupled differential equations, i.e.,
\begin{eqnarray}
  \frac{\mathrm{d}\zeta}{\mathrm{d}r} &=& \frac{\eta}{\mathcal{P}}, \label{Eq:d-e-1}\\
  \frac{\mathrm{d}\eta}{\mathrm{d}r} &=& -(\omega^2\mathcal{W}+\mathcal{Q})\zeta. \label{Eq:d-e-2}
\end{eqnarray}
To solve these oscillation equations, we adopt the boundary conditions where the center ($r=0$) of a star is static and the Lagrangian perturbation at the surface ($r=R$) must vanish, i.e.,
\begin{equation}
\delta r(0)=0,\  \eta(R)=0. \label{Eq:boundary}
\end{equation}
Then by applying Taylor expansion to $\zeta$ and $\eta$ at $r=0$ and keeping the first-order terms, we get $3\zeta_{0}=\eta_{0} /\mathcal{P}_0$, where $\zeta_0$, $\eta_0$, and $\mathcal{P}_0$ are the corresponding values of $\zeta$, $\eta$, and $\mathcal{P}$ at $r=0$. Setting $\eta_{0}=1$, we then have $\zeta_0=1/3\mathcal{P}_0$. The boundary conditions in Eq.~\eqref{Eq:boundary} correspond to the Sturm-Liouville eigenvalue problem with $\omega^2_{n}$ being the eigenvalue, which forms an infinite discrete sequence $\omega^2_{0}<\omega^2_{1}<\omega^2_{2}...$ with $\omega^2_{0}$ being the eigenfrequency of fundamental (f) mode. Note that a star is dynamically stable if the radial oscillation frequency $\omega^2_{0}>0$, which becomes unstable once $\omega^2_{0}<0$.

For a hybrid dwarf covered with normal matter, we must consider the boundary conditions at the interface of compact dwarf matter and normal matter. Since there exists Coulomb barriers between nuclei and quark nuggets, the characteristic reaction time of the transitions between nuclei and quark nuggets is much longer than that of the oscillations~\cite{Pereira2018_APJ860-12}. In such cases, the matching condition of the interface between compact dwarf matter and normal matter reads
 \begin{equation}
 [\zeta ]^+_-=0,\  [\eta ]^+_-=0,  \label{eq:slowcv}
 \end{equation}
where $[x]^+_- \equiv x^+-x^-$ with $x^+$ and $x^-$ respectively describe the properties of two types of matter above and below the interface.
\begin{figure}
\includegraphics[width=\linewidth]{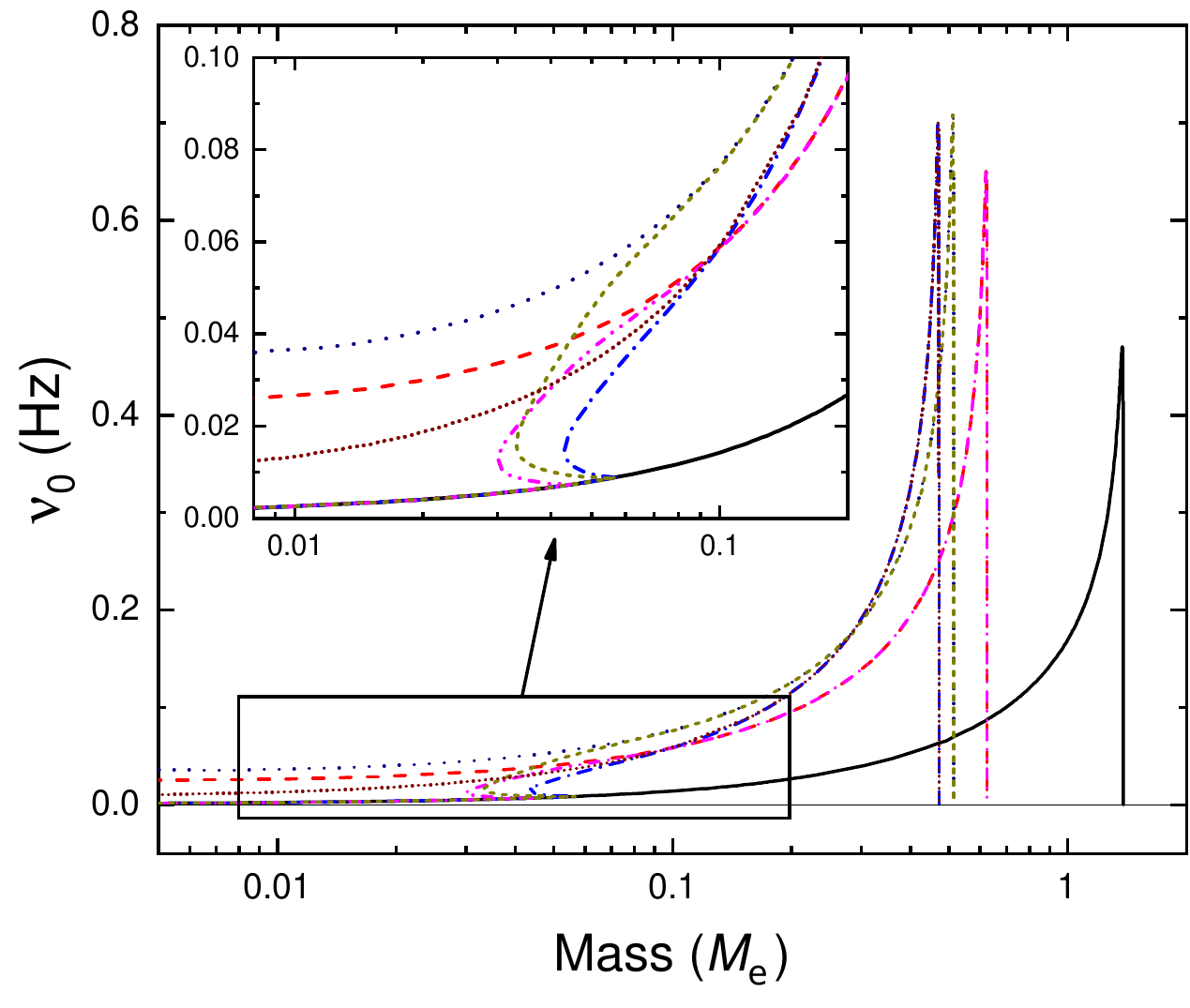}
\caption{\label{Fig:M-v0} Radial oscillation frequencies of the fundamental mode $\nu_0$ as functions of the masses of $ud$QM dwarfs and strangelet dwarfs, which correspond to the compact dwarfs indicated in the right panel of Fig.~\ref{Fig:MR}.}
\end{figure}

We first examine the radial oscillation frequencies $\nu_0$ of the fundamental mode in compact dwarfs. In Fig.~\ref{Fig:M-v0} we present the radial oscillation frequencies of the fundamental mode $\nu_0$ as functions of mass of the compact dwarfs indicated in the right panel of Fig.~\ref{Fig:MR}. In comparison with normal white dwarfs, $ud$QM dwarfs and strangelet dwarfs exhibit significantly higher frequencies since they are generally more compact than white dwarfs, which is attributed to their unique EOSs. In general, the frequency $\nu_0$ increases with the stellar mass $M$. Upon reaching its peak, $\nu_0$ quickly declines with $M$ and ultimately drops to zero at maximum mass. This aligns well with the stability criteria outlined for stellar radial oscillations, i.e., $\mathrm{d}M/\mathrm{d}E_c > 0$~\cite{Shapiro1983, Harrison1965}.

Nevertheless, for hybrid dwarfs with strangelet dwarfs and $ud$QM dwarfs enveloped in $^{16}\mathrm{O}$ white dwarf matter, the criteria $\mathrm{d}M/\mathrm{d}E_c > 0$ is no longer valid. In particular, as indicated in Fig.~\ref{Fig:MR}, the mass decreases once compact dwarf matter emerge in the center of normal white dwarfs, which finally increases with $E_c$ until reaching the maximum mass. The frequencies $\nu_0$ of hybrid dwarfs with masses ranging from 0.01 $M_\odot$ to 0.1 $M_\odot$ do not drop to zero. In fact, in this specific scenario, $\nu_0$ increases with the central density $E_c$ while the mass decrease, suggesting that the hybrid dwarfs are still stable despite $\mathrm{d}M/\mathrm{d}E_c > 0$. The reason for this is that nuclei and quark nuggets do not convert into each other due to the presence of Coulomb barriers, corresponding to the slow conversion scenarios with the boundary condition illustrated in Eq.~\eqref{eq:slowcv}. As we further increase $E_c$, the core made of strangelets or $ud$QM nuggets grows, leading to a larger $\nu_0$ in comparison with white dwarfs. Finally, at large enough $E_c$, a most massive hybrid dwarf is attained with $\nu_0\rightarrow 0$ and $\mathrm{d}M/\mathrm{d}E_c > 0$, i.e., the stability criteria still applies for most massive hybrid dwarfs. Note that for those hybrid dwarfs that are consistent with the seven ultra low-mass and small-radius white dwarfs~\cite{Kurban2022_PLB832-137204}, their oscillation periods lie within the range of 100 s to 300 s, which is similar to the observational pulsation periods of white dwarfs spanning a broad spectrum from 100 s to 7000 s~\cite{Fontaine2008_PASP120-1043}. This may offer opportunities for us to identify those exotic objects via white dwarf pulsations, in addition to the gravitational waves emitted from white dwarf binaries~\cite{Tang2023_MNRAS521-926}.

\begin{figure}
\includegraphics[width=\linewidth]{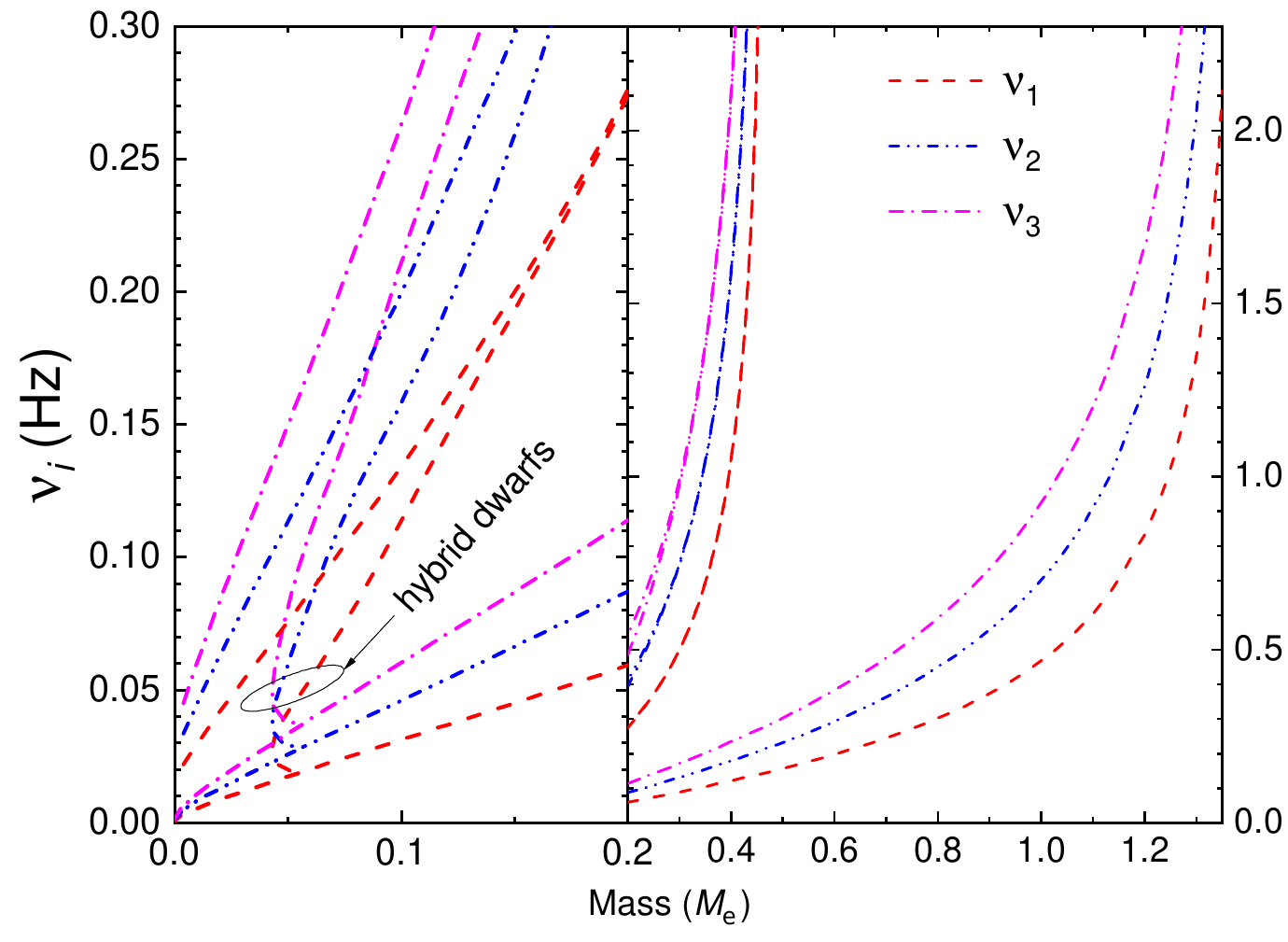}
\caption{\label{Fig:M-ff} Frequencies of the first three excited modes ($\nu_1$, $\nu_2$ and $\nu_3$) of radial oscillations as functions of the mass for $^{134}40 $ strangelet dwarfs, $^{16}\mathrm{O}$ white dwarfs, and their hybrid dwarfs.}
\end{figure}

In Fig.~\ref{Fig:M-ff}, we illustrate the frequencies of excited modes denoted as $\nu_1$, $\nu_2$, and $\nu_3$, corresponding to the node numbers 1, 2, and 3 for $^{134}40$ strangelet dwarfs, $^{16}\mathrm{O}$ white dwarfs, and their hybrid dwarfs. It is evident that the frequencies of the excited modes rise as the number of nodes increases. Unlike the fundamental mode depicted in Fig.~\ref{Fig:M-v0}, the frequencies of the excited modes do not decrease at the maximum mass but rather increase with stellar mass. The strangelet dwarfs exhibit higher excited mode frequencies compared with those of $^{16}\mathrm{O}$ white dwarfs. For hybrid dwarfs, the frequencies start to deviate from those of normal white dwarfs and approach to those of strangelet dwarfs once strangelets emerge in the core region, forming a connection between pure $^{134}40$ strangelet dwarfs and $^{16}\mathrm{O}$ white dwarfs, which is similar to those of the fundamental modes in Fig.~\ref{Fig:M-v0} and mass-radius relations in Fig.~\ref{Fig:MR}.

\section{\label{sec:con}Conclusion}
In this work, adopting an equivparticle model with both linear confinement and leading-order perturbative interactions, we study the strangelets and $ud$QM nuggets at various baryon numbers $A$ and charge numbers $Z$ at vanishing temperatures. Adopting MFA and no-sea approximations, we obtain the properties of strangelets and $ud$QM nuggets by solving the Dirac equations. Taking the parameter set $C=0.1$ and $\sqrt{D}=150$ MeV, we find that the energy per baryon of strangelets and $ud$QM nuggets decreases as baryon number $A$ increases. Both of them exhibit shell effects, which aligns with the conclusions drawn from our previous studies~\cite{Xia2018_PRD98-034031, Xia2022_PRD106-034016,You2024_PRD109-034003}. For strangelets, when $A \geq 120$, the energy per baryon is less than 930 MeV so that they are more stable than $^{56}\mathrm{Fe}$. For $ud$QM nuggets, they are unstable and may decay into nuclei if $A \leq 273$~\cite{Holdom2018_PRL120-222001}, while $ud$QM nuggets at $274\leq A \leq 446$ are more stable than the nuclei according to the liquid-type model. At $A \geq 447$, the energy per baryon of $ud$QM nuggets drops below 930 MeV. Note that the critical baryon numbers for quark nuggets to be stable as well as the shell structures may vary if we adopt different parameter sets.

We then examine the structures of compact dwarfs made of light strangelets or $ud$QM nuggets forming body-centered cubic lattices in a uniform electron background. Despite the strangelets and $ud$QM nuggets generally become more stable at larger $A$, the compact dwarfs are still stable since the fusion reactions between those objects do not take place in the presence of a Coulomb barrier $E_\mathrm{C}$, where the temperature $T\ll E_\mathrm{C}$. Then as long as  stranglets or $ud$QM nuggets do not decay into nuclei, compact dwarfs are stable if electron capture or $\beta$-decay reactions do not take place. Since the charge-to-mass ratios of strangelets and $ud$QM nuggets are smaller than that of atomic nuclei, the corresponding EOSs of compact dwarf matter become soft. Then strangelet dwarfs and $ud$QM dwarfs have much smaller masses and radii than typical white dwarfs. When the surface of a strangelet dwarf or $ud$QM dwarf is covered by a layer of normal white dwarf matter, i.e., hybrid dwarfs, their radii increase but do not exceed those of white dwarfs.

In order to examine the stability of those compact dwarfs, we analyze the frequencies of their radial oscillations. The compact dwarfs becomes unstable with the frequencies of the fundamental mode $\nu_0\rightarrow 0$, which coincide with the stability criteria $\mathrm{d}M/\mathrm{d}E_c > 0$. Nevertheless, this criteria does not apply for hybrid dwarfs, which are still stable with $\nu_0> 0$ but $\mathrm{d}M/\mathrm{d}E_c < 0$ if there exist a small core made of strangelets or $ud$QM nuggets. The frequencies of both the fundamental mode and excited modes of compact dwarfs are larger than those of ordinary white dwarfs, which may offer opportunities to identify those exotic objects. A more detailed investigation on compact dwarf oscillations are thus necessary, considering more realistic cases with non-radial oscillations and finite temperature effects with possible crystallization.

\section*{ACKNOWLEDGMENTS}
The authors would like to thank Prof. Yongfeng Huang, Prof. Sophia Han, Prof. Lap-Ming Lin, Mr. Chun-Ming Yip and Dr. Chen Zhang for fruitful discussions. This work was supported by the National Natural Science Foundation of China (Grant No. 12275234) and the National SKA Program of China (Grant Nos. 2020SKA0120300 and 2020SKA0120100).

\bibliography{yvcai}

\end{document}